\documentclass{emulateapj}

\shorttitle{Updated Optical to IR Analysis of the Virgo Cloud \ion{H}{1} 1225$+$01}
\shortauthors{Matsuoka et al.}

\begin{document}


\title{Updated Analysis of a ``Dark'' Galaxy and its Blue Companion \\in the Virgo Cloud \ion{H}{1} 1225$+$01}


\author{Y. Matsuoka\altaffilmark{1}, N. Ienaka\altaffilmark{2}, S. Oyabu\altaffilmark{1}, 
  K. Wada\altaffilmark{3}, and S. Takino\altaffilmark{4}}


\altaffiltext{1}{Graduate School of Science, Nagoya University, Furo-cho, Chikusa-ku, Nagoya 464-8602, Japan;
matsuoka@a.phys.nagoya-u.ac.jp}
\altaffiltext{2}{Institute of Astronomy, The University of Tokyo, Osawa 2-21-1, Mitaka, Tokyo 181-0015, Japan}
\altaffiltext{3}{Department of Earth and Space Science, Osaka University, Osaka 560-0043, Japan}
\altaffiltext{4}{Solar-Terrestrial Environment Laboratory, Nagoya University, Nagoya 464-8601, Japan}


\begin{abstract}
\ion{H}{1} 1225$+$01 is an intergalactic gas cloud located on the outskirts of Virgo cluster.
Its main components are two large clumps of comparable \ion{H}{1} masses ($M_{\rm HI} \sim 10^9 M_\odot$) separated by about 100 kpc.
One of the clumps hosts a blue low-surface-brightness galaxy $J$1227$+$0136, while the other has no identified 
stellar emission and is sometimes referred to as a promising candidate of a ``dark galaxy'', an optically invisible massive
intergalactic system.
We present a deep optical image covering the whole \ion{H}{1} 1225$+$01 structure for the first time, as well as 
a collection of archival data from ultraviolet to far-infrared (IR) spectral region of the brightest knot ``R1'' in 
$J$1227$+$0136.
We find that R1 has a young stellar population of age 10--100 Myr and mass $\sim 10^6 M_\odot$,
near-IR excess brightness which may point to the presence of hot dust with color temperature $\sim 600$ K, and relatively faint 
mid- to far-IR fluxes corresponding to the dust mass of up to $\sim 100 M_\odot$.
Overall, it seems to share the general properties with low-metallicity blue compact dwarf galaxies.
On the other hand, no optical counterpart to the other clump is found in our deepest-ever image.
Now the limiting surface brightness reaches down to $R_{\rm AB} > 28$ mag arcsec$^{-2}$ for any emission extended over 10\arcsec\ 
(comparable to R1), which is more than one hundred times fainter than the brightest part of the companion galaxy $J$1227$+$0136.
\end{abstract}





\section{Introduction}

The intergalactic gas cloud \ion{H}{1} 1225$+$01 was discovered serendipitously in early 1989 during routine 21 cm line observations 
at the Arecibo 305 m telescope \citep{giovanelli89}.
It was found at a supposedly blank patch of sky, toward a location which is on the outskirts of Virgo cluster and is close to the supergalactic plane.
The systematic redshift was precisely measured ($\sim$1280 km s$^{-1}$), but the distance of the cloud is uncertain due to 
its non-trivial infall motion into Virgo.
\citet{giovanelli91} proposed it is away by 0.7--1.5 times the distance of Virgo ($\sim$16 Mpc), depending on model assumptions.
\ion{H}{1} 1225$+$01 has an extended structure across $\sim 35$\arcmin, corresponding to the physical scale of $200 d_{20}$ kpc 
where $d_{20}$ is the distance of the cloud in units of 20 Mpc.
It consists of two (NE and SW) clumps separated by 17\arcmin, or $100 d_{20}$ kpc, and a bridge component 
connecting them (see Figure \ref{image}).
A subsequent Very Large Array (VLA) observation \citep{chengalur95} showed that the two clumps have similar peak column densities of 
about 1 $\times$ 10$^{21}$ atoms cm$^{-2}$.
The NE clump has a larger \ion{H}{1} mass of 2.4 $\times$ 10$^9$ $d_{20} M_{\odot}$, which accounts for a half of the total \ion{H}{1}
mass in \ion{H}{1} 1225$+$01.
The SW clump has 1.3 $\times$ 10$^9$ $d_{20} M_{\odot}$, and the rest of the mass is contained in the bridge and surrounding faint components.

In spite of the similarity in \ion{H}{1} emission features, the two clumps have a crucial difference in their optical counterparts (OCs).
Shortly after the discovery of \ion{H}{1} 1225$+$01, the OC to the NE clump was identified independently by \citet{djorgovski90},
\citet{impey90}, and \citet{mcmahon90} on the Second Palomar Sky Survey and the United Kingdom Schmidt Telescope plates.
\citet{salzer91} presented a detailed optical imaging and spectroscopic study and confirmed that the OC is a blue irregular
galaxy with an \ion{H}{2}-region like spectrum.
The measured oxygen abundance (12 $+$ log (O/H) = $7.66 \pm 0.03$) is less than a tenth of the solar value.
We call it $J$1227$+$0136 (simply from its J2000.0 coordinate) throughout this paper.
The galaxy is very faint, with the central $B$-band surface brightness $\mu_B \sim 24$ mag arcsec$^{-2}$,
and is indeed reported as one of the $\sim 700$ local low-surface-brightness (LSB) galaxies found in $\sim 800$ deg$^2$ around 
the celestial equator \citep{impey96}.
On the other hand, no OC to the SW clump has been identified to date.
The latest attempt in search was presented by \citet{turner97}, who reached down to 27.2 mag arcsec$^{-2}$
by modal filtering an $I$-band image obtained by 4.6 hr exposure with the Nickel 1.0 m telescope at Lick observatory.
Although the achieved depth is about 4 mag arcsec$^{-2}$ fainter than the brightest part of $J$1227$+$0136, 
they failed to detect any optical emission associated with the SW clump.
Thus it represents a suggestive candidate of a ``dark galaxy'', which may be an optically invisible precursor of a dwarf galaxy 
as suggested by \citet{salzer91}.

It is still not clear whether {\it truly} dark galaxies, intergalactic large gas clouds without any stellar emission, are existent or not.
There are both positive and negative predictions from theoretical models \citep[e.g.,][]{verde02,taylor05}.
Their contribution in number and mass to the local Universe provides vital information for structure formation models based on the
$\Lambda$CDM theory, which are usually tested with (intrinsically) optically visible galaxies \citep[e.g.,][]{matsuoka10,matsuoka11},
active galactic nuclei \citep[e.g.,][]{matsuoka08,matsuoka12}, and others.
Based on the past observations, \citet{briggs90} showed that dark galaxies should be extremely rare; space density of objects like 
\ion{H}{1} 1225$+$01 is estimated to be a factor of one hundred less than that of galaxies of comparable \ion{H}{1} mass.
Hunting for such a population is one of the scientific drivers in latest blind \ion{H}{1} surveys such as the \ion{H}{1} Parkes All Sky Survey
(HIPASS) and the Arecibo Legacy Fast Arecibo $L$-band Feed Array (ALFALFA) survey.
Although several attempts have been made \citep[e.g.,][]{doyle05,haynes11}, no secure identification of dark galaxies has been reported yet.
In this regard, we note that some of these searches focus on ``isolated'' dark galaxies without any dynamical companions and do not
target clouds like \ion{H}{1} 1225$+$01.
Nonetheless, the SW clump is usually referred to as the current best example of a dark galaxy candidate,
therefore its follow-up studies are of great importance as a benchmark for future surveys.
Investigations of the companion galaxy $J$1227$+$0136 as well as any optical emission throughout the gas structure are also crucial 
to resolve the nature of \ion{H}{1} 1225$+$01 as a whole.

In this paper, we present an updated analysis of the two clumps by exploiting an original deep optical image and a collection of archival data 
from ultraviolet (UV) to far-infrared (IR) spectral region.
Although there are other two clumps, one called the bridge and another to the NNW of the NE clump by $\sim$5\arcmin, they have not been 
subjects of intensive studies historically.
It is simply because of their less \ion{H}{1} mass and lower peak column density by a factor of a few than the main (NE and SW) clumps,
and their proximity to the largest (NE) clump which may indicate that they are more likely tidal features of the NE clump.
However, we do not intend to deny the possibility that the smaller clumps are intrinsically isolated systems and can be regarded 
as precursors of dwarf galaxies.

All magnitudes measured in this work are presented on the AB system, while the $B$- and $I$-band magnitudes taken from the literature
are based on the Vega system.
The distance of \ion{H}{1} 1225$+$01 is $d_{20}$ in units of 20 Mpc.

\section{Observation}

The original deep optical image was obtained as part of our ongoing project \citep{matsuoka11b} to search for 1 $\micron$ excess sources 
in the United Kingdom Infrared Telescope (UKIRT) Infrared Deep Sky Survey \citep[UKIDSS;][]{lawrence07}.
We used a wide-field mosaic CCD camera MOA-cam3 mounted at the prime focus of the Microlensing Observations in Astrophysics (MOA) 1.8 m 
telescope \citep{sumi11}.
The telescope has been build on Mt. John Observatory, New Zealand, by Nagoya University in collaboration with the University of Canterbury.
MOA-cam3 is equipped with ten 2k $\times$ 4k CCD chips and provides very wide field of view of 2.18 deg$^2$ with 0\arcsec.58 pixel scale \citep{sako08}.
Our project utilizes a special $R$-band filter characterized by broad spectral coverage from $\lambda$ = 6200 \AA\ to 8700 \AA\ with 
the similar effective wavelength to that of the Sloan Digital Sky Survey \citep[SDSS;][]{york00} $i$-band filter.

The observations of the \ion{H}{1} 1225$+$01 field were carried out on nine nights in 2012 January and February.
The average seeing was 2\arcsec.9 with a significant night-to-night variation (up to $\sim$5\arcsec).
The poor seeing condition is acceptable in this work focusing on extended emission with relatively large ($>$5\arcsec) apparent sizes.
We obtained 83 shots of 5 minute exposure, for total integration of 6.9 hr.
In order to minimize the effects of large scale non-uniformity of the detector sensitivity and fringing, we configured the telescope dithering pattern
so that a given field was observed on several different CCD chips.

Data reduction was performed in a standard manner.
Dark current signals have been measured on each night of the observations, and were subtracted from the science frames.
Flat fielding was achieved by using dome flat images.
Fringing patterns of the CCD chips were estimated by stacking flat-fielded science frames without offsetting (that would compensate for the telescope 
dithering), and removed.
Finally the science frames were stacked into a final deep image, after manually removing artifacts such as satellite trails.
Photometric calibration was achieved by referring to SDSS $i$-band magnitudes of bright (15--18 mag) stars within the observed field of view.
Thanks to the similarity in effective wavelengths of the MOA $R$ and SDSS $i$ filters, we found a very good linear correlation between
the two sets of magnitudes with less than 0.1 mag scatter (root mean square).

\section{Results and Discussion}

\begin{figure*}
\epsscale{2.2}
\includegraphics[angle=90,scale=0.9]{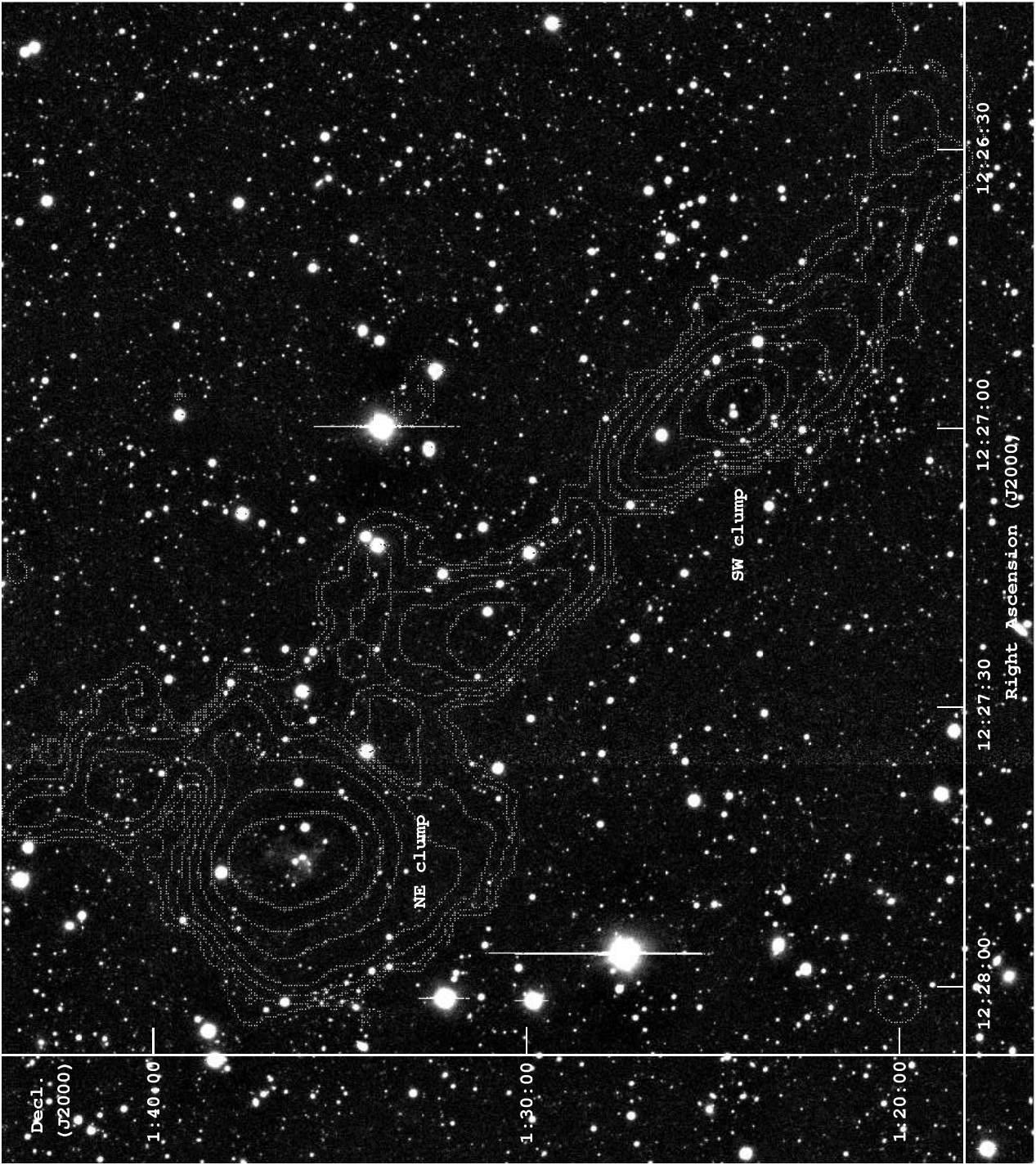}
\caption{MOA $R$-band image of the \ion{H}{1} 1225$+$01 field (31\arcmin $\times$ 28\arcmin).
  The contours represent the VLA \ion{H}{1} emission map and are logarithmically spaced at 5.0, 11.0, 20.5, 41.6, 84.2, 170.1, 
  and $345.6 \times 10^{18}$ atoms cm$^{-2}$.
  The clean beam size of the VLA observation is shown by the dotted circle at the southeast corner \citep{chengalur95}.
\label{image}}
\end{figure*}

The reduced image of the \ion{H}{1} 1225$+$01 field (31\arcmin $\times$ 28\arcmin) is presented in Figure \ref{image}.
On top of the stellar (and nebular) distribution, we show the VLA \ion{H}{1} emission map obtained by \citet{chengalur95}.
Our data cover the whole \ion{H}{1} structure and provide, to our knowledge, the deepest-ever image of the field.
We first derive some physical properties of the OC to the NE clump, $J$1227$+$0136, and then present the results on
a search for optical emission from other regions.

\subsection{NE clump}

\begin{figure}
\epsscale{1.0}
\plotone{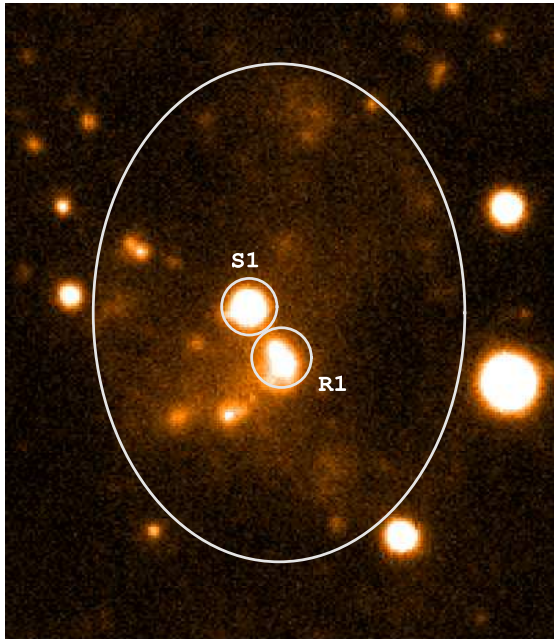}
\caption{Close-up view of $J$1227$+$0136, the OC to the NE clump of \ion{H}{1} 1225$+$01.
  The large ellipse (140\arcsec $\times$ 100\arcsec) represents an aperture used for the photometry, in which a foreground star ``S1'' is excluded.
  The brightest knot ``R1'' and the 12\arcsec\ circular aperture on it are also shown (see the text).
  \label{NEcomp}}
\end{figure}

The galaxy $J$1227$+$0136 is clearly detected at the \ion{H}{1} peak position of the NE clump. 
Its close-up view is given in Figure \ref{NEcomp}.
As is previously known, the galaxy has a highly irregular morphology with the brightest knot, which we call ``R1'' hereafter, at close to
the galactic center.
The J2000.0 coordinate of R1 is R.A. 12$^{\rm h}$27$^{\rm m}$46$^{\rm s}$.1, decl. +01$^{\circ}$36$^{\rm m}$01$^{\rm s}$.
We carried out photometry of the galaxy within an elliptical aperture of 140\arcsec $\times$ 100\arcsec,
so that most of the visible diffuse light is collected and contamination from surrounding sources is minimized.
A point source ``S1'' marked in Figure \ref{NEcomp} is excluded since its much redder color than the rest of the emission in the 
aperture indicates it is a foreground star \citep{salzer91}.
We estimated the underlying sky brightness and its uncertainty from mean and scatter of the background levels measured at several, 
relatively blank fields outside the aperture.
The obtained total and surface brightness are 16.02 $\pm$ 0.04 mag and 25.51 $\pm$ 0.04 mag arcsec$^{-2}$, respectively.
We also measured brightness of R1 within a circular 12\arcsec\ aperture,
which results in 17.93 $\pm$ 0.01 mag and 23.08 $\pm$ 0.01 mag arcsec$^{-2}$.
The presented magnitudes have been corrected for the Galactic extinction assuming $A_V = 0.062$ mag at the line-of-sight 
to the galaxy \citep{sfd98} and the extinction curve presented by \citet{pei92}.

$J$1227$+$0136 has also been imaged by the SDSS\footnote{
Funding for the SDSS and SDSS-II has been provided by the Alfred P. Sloan Foundation, the Participating Institutions, the National Science Foundation, 
the U.S. Department of Energy, the National Aeronautics and Space Administration, the Japanese Monbukagakusho, the Max Planck Society, and the Higher 
Education Funding Council for England. The SDSS Web Site is http://www.sdss.org/.
The SDSS is managed by the Astrophysical Research Consortium for the Participating Institutions. The Participating Institutions are the American Museum 
of Natural History, Astrophysical Institute Potsdam, University of Basel, University of Cambridge, Case Western Reserve University, University of Chicago, 
Drexel University, Fermilab, the Institute for Advanced Study, the Japan Participation Group, Johns Hopkins University, the Joint Institute for Nuclear 
Astrophysics, the Kavli Institute for Particle Astrophysics and Cosmology, the Korean Scientist Group, the Chinese Academy of Sciences (LAMOST), 
Los Alamos National Laboratory, the Max-Planck-Institute for Astronomy (MPIA), the Max-Planck-Institute for Astrophysics (MPA), New Mexico State University, 
Ohio State University, University of Pittsburgh, University of Portsmouth, Princeton University, the United States Naval Observatory, and the University 
of Washington.} 
and the UKIDSS\footnote{
This work is based in part on data obtained as part of the UKIRT Infrared Deep Sky Survey.} 
Large Area Survey.
We inspected the images retrieved from the archives and found that, although they are much shallower than the MOA $R$ image, the brightest 
knot R1 is visible in all but UKIDSS $K$ band data.
The measured magnitudes of R1 within the circular 12\arcsec\ aperture are summarized in Table \ref{tab:photo}.
The Galactic extinction has been corrected as above.
In addition, we found that the galaxy was observed in 2006 June and July by the {\it Spitzer Space Telescope}\footnote{
This work is based in part on observations made with the {\it Spitzer Space Telescope}, obtained from the 
NASA/ IPAC Infrared Science Archive, both of which are operated by the Jet Propulsion Laboratory, 
California Institute of Technology under a contract with the National Aeronautics and Space Administration.
} ({\it Spitzer})
with the Infrared Array Camera \citep[IRAC;][]{fazio04} and the Multiband Imaging Photometer for {\it Spitzer} \citep[MIPS;][]{rieke04}.
The AOR keys are 17567232 and 17574144 for the IRAC and MIPS observations, respectively.
R1 is detected only in the two shorter-wavelength IRAC bands (3.6 and 4.5 $\mu$m), in which the brightness was measured as listed in Table \ref{tab:photo}.
We provide 3$\sigma$ upper limits of flux in IRAC 5.8, 8.0 $\mu$m and MIPS 24, 70 $\mu$m bands where no robust detection was confirmed.
We used the 12\arcsec\ aperture in all but the MIPS 70 $\mu$m band, in which an alternative 32\arcsec\ aperture was used in consideration of 
the relatively poor angular resolution.
Appropriate aperture corrections were applied to these {\it Spitzer} magnitudes by referring to the IRAC and MIPS Handbooks.


\begin{table}
  \begin{center}
    \caption{Photometry of the Brightest Knot ``R1'' in $J$1227$+$0136 \label{tab:photo}}
    \begin{tabular}{cccc}
      \tableline\tableline
             &      & Magnitude & Flux \\
             & Band & (AB mag)  & (mJy)\\
      \tableline
      \tableline
      SDSS   & $u$        & 18.14 $\pm$ 0.06 & 0.20 $\pm$ 0.01 \\
             & $g$        & 17.89 $\pm$ 0.06 & 0.25 $\pm$ 0.01 \\
             & $r$        & 17.89 $\pm$ 0.06 & 0.25 $\pm$ 0.01 \\
             & $i$        & 18.11 $\pm$ 0.09 & 0.21 $\pm$ 0.02 \\
             & $z$        & 18.22 $\pm$ 0.23 & 0.19 $\pm$ 0.04 \\
      UKIDSS & $Y$        & 18.16 $\pm$ 0.12 & 0.20 $\pm$ 0.02 \\
             & $J$        & 18.15 $\pm$ 0.23 & 0.20 $\pm$ 0.04 \\
             & $H$        & 18.29 $\pm$ 0.26 & 0.18 $\pm$ 0.04 \\
             & $K$        & $>$ 17.8         & $<$ 0.27    \\
      IRAC   & 3.6 $\micron$ & 18.97 $\pm$ 0.07 & 0.094 $\pm$ 0.006 \\
             & 4.5 $\micron$ & 19.00 $\pm$ 0.07 & 0.091 $\pm$ 0.006 \\
             & 5.8 $\micron$ & $>$ 18.5         & $<$ 0.14    \\
             & 8.0 $\micron$ & $>$ 18.4         & $<$ 0.16    \\
      MIPS   & 24  $\micron$ & $>$ 17.0         & $<$ 0.60    \\
             & 70  $\micron$ & $>$ 12.2         & $<$ 48      \\
      \tableline
    \end{tabular}
  \end{center}
\end{table}

\begin{figure}
\epsscale{1.0}
\plotone{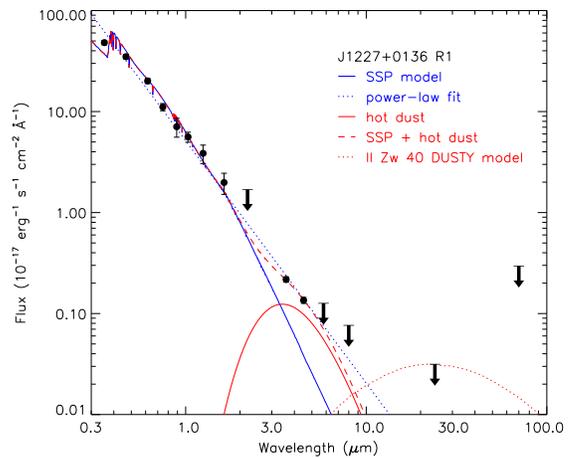}
\caption{Broadband SED of the brightest knot ``R1'' in $J$1227$+$0136, based on the SDSS, UKIDSS, and {\it Spitzer} magnitudes (dots and arrows;
  the latter represent 3$\sigma$ upper limits).
  Several model SEDs are shown by the lines; the power-law fit $F_\lambda \propto \lambda^{-2.4}$ (blue dotted);
  the best-fit SPS model (blue solid); 
  hot dust represented by a modified blackbody with $T$ = 600 K and $\lambda^{-2}$ emissivity (red solid); 
  the sum of the SSP and the hot dust emissions (red dashed); 
  and the DUSTY model of II Zw 40 normalized to the observed upper limit in the MIPS 24 $\mu$m band (red dotted).
  \label{NEsed}}
\end{figure}

It is evident from Figure \ref{NEsed} that $J$1227$+$0136 R1 has a very blue spectral energy distribution (SED).
The 0.3--5.0 $\mu$m SED is well described by a single power-law function, $F_\lambda \propto \lambda^{-2.4}$.
We estimated its stellar properties by comparing the SDSS and UKIDSS magnitudes
to a stellar population synthesis (SPS) model provided by the \texttt{GALAXEV} \citep{bc03} code.
A simple stellar population, in which all stars are formed in an instantaneous starburst, is assumed in the SPS calculations.
We adopt two values of metallicity, $Z$ = 0.02 $Z_{\odot}$ and 0.2 $Z_{\odot}$, in consideration of the gas-phase metallicity $\sim$0.07 $Z_{\odot}$ 
found by \citet{salzer91}.
The age grid was $t_{\rm age}$ = 0.001, 0.003, 0.01, 0.03, 0.1, 0.3, 1, 3, and 10 Gyr.
The fitting results based on the least-$\chi^2$ method are summarized in Table \ref{tab:sps}.
The best-fit parameters without dust extinction are $Z$ = 0.02 $Z_{\odot}$, $t_{\rm age}$ = 100 Myr, and the stellar mass 
$M_{\rm star} = 3.5 \times 10^{6} d_{20}^2\ M_{\odot}$.
The reduced $\chi^2$ value, $\chi^2_{\rm red} = 1.5$, indicates a reasonable agreement between the observed and model SEDs.
We show the best-fit SPS model in Figure \ref{NEsed}.
The higher metallicity $Z$ = 0.2 $Z_\odot$ results in a slightly worse fit and younger age solution ($t_{\rm age}$ = 10 Myr),
due to the notorious age-metallicity degeneracy.
When the effect of dust extinction is incorporated into the SPS calculations, assuming the \citet{calzetti00} extinction law 
and variable $A_V$ values between 0.0 and 0.5 mag, 
we obtain $t_{\rm age}$ = 10--30 Myr with the derived mass $M_{\rm star} = (0.5-2.5) \times 10^{6} d_{20}^2\ M_{\odot}$.
Therefore, we conservatively conclude that R1 has stellar age of 10--100 Myr and mass of $\sim 10^6 d_{20}^2\ M_\odot$.
The age is consistent with the result of \citet{salzer91} who derived $t_{\rm age} < 40$ Myr for the diffuse LSB portions (not R1) of the galaxy.
The whole galaxy seems to contain 10 times larger mass ($\sim 10^7 d_{20}^2\ M_\odot$) based on their analysis.

\begin{table*}
  \begin{center}
    \caption{Stellar Population Properties of $J$1227$+$0136 R1 \label{tab:sps}}
    \begin{tabular}{ccccccc}
      \tableline\tableline
Dust         & Metallicity & Age   & $A_V$ &  Mass                   & \\
Extinction?    & ($Z_\odot$) & (Myr) & (mag) & ($10^{6} d_{20}^2\ M_{\odot}$) & $\chi^2_{\rm red}$ ($\chi^2$) \\
\tableline
No  & 0.02   & 100   & \nodata & 3.5 & 1.5 (10.7)\\
    &  0.2   &  10   & \nodata & 0.5 & 1.8 (12.9)\\
\tableline
Yes & 0.02   &  30   & 0.4     & 2.5 & 0.8  (5.9)\\
    &  0.2   &  10   & 0.0     & 0.5 & 1.8 (12.9)\\
      \tableline
    \end{tabular}
  \end{center}
\end{table*}

At near-IR wavelengths beyond the UKIDSS bands, the observed SED becomes apparently red compared to the stellar emission model:
the predicted and observed [3.6]$-$[4.5] colors are $-$0.5 and 0.0 mag (AB), respectively. 
The presence of such a red excess component in galaxies has been known and discussed for the last several years.
\citet{lu03} showed that the excess, found in a sample of star-forming disk galaxies, is best explained by hot dust emission with color
temperature $T$ = 750--1000 K, rather than dust reddening of stellar continuum.
They suggested that this emission may be the same component that is seen in reflection nebulae and the large-scale interstellar medium
of the Milky Way.
Hot dust in dwarf and irregular galaxies has also been indicated and studied by mostly exploiting the near- to mid-IR imaging capability 
of {\it Spitzer} \citep[e.g.,][]{engelbracht05,hunter06}.
We found that the excess brightness of $J$1227$+$0136 R1 at 3.6 and 4.5 $\micron$ points to the color temperature of $T \sim 600$ K (a modified
blackbody with $\lambda^{-2}$ emissivity is assumed; see Figure \ref{NEsed}), in agreement with the hot dust origin for this component.
The estimated mass in this dust population is $\sim 3 \times 10^{-5} M_{\odot}$ \citep[mass absorption coefficients $\kappa_{\rm abs}$ 
were taken from][]{li01}.
Meanwhile, we note that other mechanisms such as strong Br$\alpha$ line and nebular continuum emissions are also suggested as non-negligible
contributors to these IRAC bands \citep[e.g.,][]{smith09}.

We could not detect the galaxy at beyond $\lambda = 5$ $\micron$ (up to $\sim 100$ $\micron$), where normal starburst SEDs are dominated by warm dust and 
polycyclic aromatic hydrocarbon emissions.
This (and longer wavelength) spectral region of low-metallicity blue compact dwarf galaxies (BCDs), which are most analogous to R1 in terms of 
stellar content, metallicity, and size ($\le$ 1 kpc), is being intensively studied as a testbed for star formation and dust evolution processes
in high-redshift Universe \citep[e.g.,][]{hunt05}.
While dust is produced predominantly in the envelopes of asymptotic giant branch stars in old (stellar age $>$ 1 Gyr) systems, 
low-metallicity BCDs have not evolved enough to have such a production channel, and hence Type II supernovae are thought to be the main source 
of dust in these objects.
Evidence of supernova origin for dust is also seen in larger systems such as luminous high-redshift quasars \citep[e.g.,][]{maiolino04} and young 
ultraluminous IR galaxies \citep[e.g.,][]{kawara11}.

Although a detailed analysis is hampered by the faintness of the galaxy at IR wavelengths, here we try to provide the upper limits of mass in warm dust 
($M_{\rm dust}$) 
based on the observed flux upper limits.
For this purpose, we assume the $M_{\rm dust}$ to $L_{24\micron}$ (monochromatic luminosity at 24 $\micron$) ratio found in a low-metallicity BCD 
II Zw 40. 
The dust mass in II Zw 40 was estimated by \citet{hunt05} by comparing its mid-IR to radio fluxes to the model SEDs constructed from the 
\texttt{DUSTY} \citep{ivezic99} code, which is designed to perform an appropriate radiative transfer calculation in dusty environments.
The monochromatic luminosity at 24 $\micron$ was chosen because the observed upper limit puts the strongest constraint 
on the amplitude of the model SED (see Figure \ref{NEsed}).
The derived mass is $M_{\rm dust} \la 100 d_{20}^2\ M_\odot$, thus the whole galaxy may contain up to $M_{\rm dust} \sim 1000 d_{20}^2\ M_\odot$
if we adopt the same scaling factor as for stellar content ($M_{\rm star} \sim$ $10^6$ and $10^7 d_{20}^2\ M_\odot$ for R1 and the whole galaxy,
respectively; see above).
This is rather low compared to other BCDs of comparable or smaller \ion{H}{1} masses \citep[e.g.,][]{hirashita08}, a part of which may be due to
the fact that $J$1227$+$0136 has lower metallicity than most of known BCDs.
However, the current result is based only on 
the brightness in a single mid-IR band and could be significantly altered with future far-IR observations.

\subsection{SW clump}

No diffuse emission associated with the SW clump or any \ion{H}{1} components other than the NE clump is found in the MOA $R$ image 
(Figure \ref{image}).
The 3$\sigma$ limiting surface brightness is 25.2 mag arcsec$^{-2}$, which is slightly deeper than that achieved in the previous attempt 
(only for the SW region) by \citet{turner97}.
We smoothed the image on a 10\arcsec\ $\times$ 10\arcsec\ box by taking the mode value in each of the boxes, which improved the 
limiting depth to 28.3 mag arcsec$^{-2}$, but did not find any diffuse emission as well in the relevant fields.
Therefore the OC to the SW clump, if it is present and extended over $>$10\arcsec\ (comparable to R1), should have the surface 
brightness of more than one hundred times fainter than the brightest part of the companion galaxy $J$1227$+$0136.
At the same time, we cannot reject the possibility that star formation in the SW clump is prevented/delayed for some reasons
and the clump is truly invisible in optical light.

The formation process of \ion{H}{1} 1225$+$01 and the nature of the SW clump is still not well understood.
\citet{chengalur95} concluded that the SW clump is an edge-on rotating disk which is kinematically separated from other \ion{H}{1}
components, based on its geometry, high central column density, and velocity distribution.
Indeed, its position - velocity diagram is very similar to those along the major axis of disk galaxies.
The inferred rotation velocity is rather small, $\sim$13 km s$^{-1}$ (a few times smaller than that of the NE clump).
They suggested that the entire \ion{H}{1} 1225$+$01 structure was formed via tidal interactions between the NE and SW clumps, which have
comparable \ion{H}{1} mass and density and were once dynamically distinct.
In that case, the clear contrast between the two clumps in their stellar contents may hint at some key ingredients of star formation process.
\citet{impey97} showed that, on the diagram of surface mass density and velocity dispersion, the NE clump departs significantly from 
the locus of normal dwarf/irregular galaxies toward the region in which the energy dissipation and subsequent star formation are not allowed
\citep[based on][]{efstathiou83}.
Qualitatively, at the velocity dispersion found in the NE clump, the surface brightness (corresponding to surface mass) of normal galaxies 
and the NE clump are $\mu_B \sim 24$ and 29 mag arcsec$^{-2}$, respectively, while no dissipation is allowed at $\mu_B > 30$ mag arcsec$^{-2}$ 
for primordial gas.
The SW clump has the similar \ion{H}{1} mass density and velocity dispersion to the NE counterpart,
but some small (but critical) difference may have prevented it from setting out to the dissipation process.
Alternatively, the SW clump might be an intimately associated structure with the NE clump, such as a large tidal tail created in a past merger 
event that the NE clump has undergone \citep{turner97}.

Meanwhile, some theoretical and observational studies suggest that dark galaxies cannot live long.
\citet{taylor05} found that, in the absence of an internal radiation field, the gas in a galaxy with baryonic mass greater than $10^8 M_{\odot}$
becomes gravothermally unstable via H$_2$ cooling and leads inevitably to star formation.
Their model shows that the fraction of the unstable (to star formation) gas is smaller in galaxies with lower baryonic mass.
In fact, \citet{warren07} found that the stellar to baryonic mass ratio decreases toward lower baryonic mass in $\sim$40 late-type galaxies
observed in the local Universe, which is consistent with the above prediction.
They concluded that galaxies with very high \ion{H}{1} mass to optical light ratios are extremely rare because such systems can only be produced 
with a combination of the physical conditions such as low initial mass, shallow dark matter potential, and low-density environment.
These studies seem to deny the possibility that the SW clump, located in the relatively high-density environment within the Virgo cluster, 
is a dark galaxy that has persisted and will persist for very long time.

The fact that the SW clump of \ion{H}{1} 1225$+$01 remains as (one of) the best example(s) of dark galaxy candidates, if it is not a tidal
feature of the NE clump, for more than 20 years since 
its discovery may indicate that they are in fact extremely rare, as suggested by \citet{briggs90}.
Future studies on such a population will benefit greatly from the coming deep and wide optical surveys, e.g., the planned Hyper-Suprime Cam 
survey with Subaru telescope \citep{takada10}, in combination with the existing \ion{H}{1} surveys such as the HIPASS and the ALFALFA.

\acknowledgments

We are grateful to the referee for providing the useful comments to improve the paper.
This work was supported by a Grant-in-Aid for Young Scientists (22684005) and the Global COE Program of Nagoya University 
``Quest for Fundamental Principles in the Universe'' from JSPS and MEXT of Japan.

\end{document}